\documentclass[a4paper,fleqn,usenatbib]{mnras}

\usepackage{newtxtext,newtxmath}

\usepackage[T1]{fontenc}
\usepackage{ae,aecompl}

\usepackage{graphicx}	
\usepackage {threeparttable}
\usepackage{amsmath}	
\usepackage{amssymb}	

\title[Performance test of QU-fitting]{Performance test of QU-fitting in cosmic magnetism study}

\author[Miyashita et al.]{
Yoshimitsu Miyashita,$^{1}$\thanks{170d9004@st.kumamoto-u.ac.jp}
Shinsuke Ideguchi,$^{2}$
Shouta Nakagawa,$^{1}$ 
\newauthor
Takuya Akahori,$^{3}$
and Keitaro Takahashi$^{1}$
\\
$^{1}$Kumamoto University, 2-39-1, Kurokami, Kumamoto 860-8555, Japan\\
$^{2}$Department of Physics, UNIST, Ulsan 44919, Korea\\
$^{3}$Graduate School of Science and Engineering, Kagoshima University, Kagoshima 890-0065, Japan
}

\date{Accepted XXX. Received YYY; in original form ZZZ}

\pubyear{2017}

\begin{document}
\label{firstpage}
\pagerange{\pageref{firstpage}--\pageref{lastpage}}
\maketitle

\begin{abstract}
QU-fitting is a standard model-fitting method to reconstruct distribution of magnetic fields and polarized intensity along a line of sight (LOS) from an observed polarization spectrum. In this paper, we examine the performance of QU-fitting by simulating observations of two polarized sources located along the same LOS, varying the widths of the sources and the gap between them in Faraday depth space, systematically. Markov Chain Monte Carlo (MCMC) approach is used to obtain the best-fit parameters for a fitting model, and Akaike and Bayesian Information Criteria (AIC and BIC, respectively) are adopted to select the best model from four fitting models. We find that the combination of MCMC and AIC/BIC works fairly well in model selection and estimation of model parameters in the cases where two sources have relatively small widths and a larger gap in Faraday depth space. On the other hand, when two sources have large width in Faraday depth space, MCMC chain tends to be trapped in a local maximum so that AIC/BIC cannot select a correct model. We discuss the causes and the tendency of the failure of QU-fitting and suggest a way to improve it.
\end{abstract}




\section{Introduction}

Faraday tomography is a sophisticated technique which allows us to probe cosmic magnetism with Faraday spectrum \citep{bur66,bre05}. Faraday spectrum represents the distribution of polarized intensity as a function of Faraday depth, which is proportional to an integration of thermal electron density and magnetic fields along the line of sight (LOS). Compared with the conventional Faraday rotation measure technique, Faraday spectrum gives us much richer information on LOS distributions of thermal and cosmic-ray electron densities and magnetic fields. Therefore, Faraday tomography is expected to become a new transformational technique in polarimetric radio-astronomy.

Faraday tomography has been applied to various observations in the interstellar medium \citep{Schnit07,Schnit09}, galactic magnetic fields \citep{bec09a,hea09,govo10,mao10,woll10},  and active galactic nuclei \citep{osu12}. However, in order to utilize this technique, we are confronted with two technical problems. One is the interpretation of Faraday spectrum. In general, Faraday spectrum has a very complicated shape with a lot of spikes \citep{bell11,fri11,beck12,ide14b}. Since Faraday depth does not generally have one-to-one correspondence with the physical distance, it is not easy to understand spatial structures of physical quantities along the LOS. \citet{ide17} suggested to use some statistical quantities to extract global features of galactic magnetic fields.

The other problem is the reconstruction of Faraday spectrum from observed polarized intensity and this is the focus of this paper. Because Faraday tomography is based on Fourier transformation in frequency domain, frequency coverage of the observation is a primary parameter which determines the quality of reconstruction \citep{aka14}. We can obtain only a finite frequency coverage of polarized emission in real observation, resulting in imperfect reconstruction of Faraday spectrum. In order to improve the quality, several techniques have been proposed. For example, RM CLEAN which removes the siderobes of the dirty Faraday spectrum \citep{hog74,hea09,kum14,miya16}, QU-fitting which is a model fitting method \citep{osu12,ide14a,ozawa,kac17}, and compressive sensing which assumes a sparsity of Faraday spectrum \citep{li11a,li11b,and12} have been widely used. 

Recently, \citet{sun15} held a benchmark test to evaluate capabilities of these techniques. They reported that QU-fitting exhibits better results in many cases. Nevertheless, detailed test of capabilities of QU-fitting has not been done. Therefore, this paper examine the performance of QU-fitting in a more systematic manner.

In QU-fitting, we need to explore the parameter space of a given fitting model and need to find the best-fit parameter set. In such a parameter search, Markov Chain Monte Carlo (MCMC) approach is known to be very efficient. Actually, some authors adopted this technique in the benchmark test. However, it is also known that MCMC suffers from the so-called ``local maximum problem", where MCMC chain is trapped in a local maximum of likelihood function and cannot reach the best-fit parameter set. We consider various source models of Faraday spectrum to examine this problem in Faraday tomography.

Another important ingredient of Faraday tomography is model selection. Because we do not know the correct model a priori, we need to try several plausible fitting models and select the best one. In this situation, Akaike Information Criterion (AIC) and Bayesian Information Criterion (BIC) are often used. They quantify the balance between the fit to data and the simplicity of the model. We investigate the effectiveness of these criteria as well.

In this study, we evaluate the capability of QU-fitting through a series of simulations which consist of making mock observation data, fitting with MCMC and model selection with AIC and BIC. In section 2, we explain the details of Faraday tomography, QU-fitting and the model setting. We show the results in section 3 and we discuss the reasons of failure cases and propose a way to improve in section 4. Finally, we give a summary in section 5.

\section{Model and calculation}

\subsection{Faraday Tomography}

We follow the standard formalism of Faraday tomography described in the literatures \citep{bur66,bre05}. Complex polarized intensity P($\lambda^2$) is expressed as:
\begin{equation}
P(\lambda^2) = Q(\lambda^2) + i U(\lambda^2) = \int_{-\infty}^{\infty} F(\phi) e^{2i\phi\lambda^2} d\phi,
\label{eq:Fourier}
\end{equation}
where $Q$ and $U$ are the Stokes parameters, and {\it F}($\phi$) is Faraday dispersion function (FDF) or Faraday spectrum, which is the complex polarized intensity in Faraday depth space. Faraday depth $\phi$ is defined as
\begin{equation}
\phi \approx 811.9\int n_e B_{||} dx~~~{\rm rad~m^{-2}},
\label{equ2}
\end{equation}
where $n_e$ is the number density of thermal electrons in cm$^{-3}$, $B_{||}$ is the LOS component of magnetic fields in $\mu$G, and {\it x} is the physical distance to the target source in pc.

As a general method to estimate the Faraday spectrum, we perform inverse Fourier transformation of equation (\ref{eq:Fourier}) as follows
\begin{equation}
F(\phi) = \frac{1}{\pi} \int_{-\infty}^{\infty} P(\lambda^2) e^{-2i\phi\lambda^2} d\lambda^2.
\label{equ3}
\end{equation}
We, however, cannot perform this transformation perfectly, because the observable wavelength coverage is limited. We rewrite the Faraday spectrum using a window function W($\lambda^2$), where W($\lambda^2$) = 1 if $\lambda^2$ is in the observable bands and otherwise W($\lambda^2$) = 0, as follows
\begin{eqnarray}
\label{equ4} 
\tilde{F}(\phi) & = & \frac{1}{\pi} \int_{-\infty}^{\infty} W(\lambda^2)P(\lambda^2) e^{-2i\phi\lambda^2} d\lambda^2, \\
\label{equ5}
& = & \frac{1}{K} F(\phi)*R(\phi).
\end{eqnarray}
$\tilde{F}(\phi)$ is called dirty Faraday spectrum and is incomplete spectrum because the limitation of wavelength coverage produces a siderobe such as a beam pattern in radio interferometry. We can describe the dirty Faraday spectrum as a convolution mathematically between {\it F}($\phi$) and rotation measure spread function (RMSF) {\it R}($\phi$), inverse Fourier transformation of the window function
\begin{eqnarray}
\label{equ6}
R(\phi)  & = & K \int_{-\infty}^{\infty} W(\lambda^2) e^{-2i\phi\lambda^2} d\lambda^2,\\
\label{equ7}
K^{-1} & = & \int_{-\infty}^{\infty} W(\lambda^2) d\lambda^2,
\end{eqnarray}
where $K$ is the normalization of RMSF and a shape of RMSF is like a sinc function. {\it R}($\phi$) determines an accuracy of Faraday spectrum.

\subsection{QU-fitting}

QU-fitting is a model-fitting method, where we compare observed polarized intensity, $P(\lambda^2)$, with a fitting model. A fitting model is often given as a function of $\lambda^2$, and it has a specific form and parameters based on theoretical considerations such as depolarization \citep{osu12,kac17}. Contrastingly, an FDF with parameters, $F(\phi)$, can also be a fitting model \citep{ide14a,ozawa}, while it should be Fourier-transformed into the polarized intensity to be fitted (see, Eq.($\ref{eq:Fourier}$)). Once a fitting model is given, the best-fit parameter set is sought. Because the number of parameters is relatively large ($\gtrsim 10$), MCMC method is an effective way to search for it in the parameter space.

When we have multiple fitting models and need to select the best model from them, criteria such as AIC and BIC are often used. They are calculated, for each model, by,
\begin{eqnarray}
{\rm AIC} &=& -2 \log L(\hat\theta) + 2k, \\
{\rm BIC} &=& -2 \log L(\hat\theta) + k \log n,
\end{eqnarray}
where $L(\hat\theta)$ is the model likelihood for the best-fit parameter set, $\hat\theta$ (see below), $k$ is the number of parameters, and $n$ is the number of data. The first term represents the goodness of the fit between the model and observation data, and the second term represents a penalty due to the number of parameters. Note that BIC imposes a larger penalty than AIC when $n$ is very large as the current situation ($n = 2,200$). The fitting model with the smallest value is regarded as the best model.

\subsection{Method and Models}

In this paper, we simulate the above procedures to evaluate the effectiveness of QU-fitting. To do this, we assume an FDF as a source model and make mock data of polarized intensity by Fourier-transforming the assumed FDF and adding observation noises. Then, we fit the mock data with several fitting models through MCMC and select the best model using AIC and BIC.

We consider a source model with two Gaussian functions:
\begin{eqnarray}
F(\phi) &=& \frac{f_1}{\sqrt{2\pi\sigma_1^2}} \exp{\left( -\frac{(\phi - \phi_{1})^2}{2\sigma_1^2} \right)} e^{2i\chi_{0,1}} \nonumber \\ 
&+& \frac{f_2}{\sqrt{2\pi\sigma_2^2}} \exp{\left( -\frac{(\phi - \phi_{2})^2}{2\sigma_2^2} \right)} e^{2i\chi_{0,2}}.
\label{eq:Gauss}
\end{eqnarray}
This model corresponds to two independent polarized sources or two components within a source along a single LOS. Here, $f_1$ and $f_2$ are the brightness, $\phi_1$ and $\phi_2$ are Faraday depths, $\sigma_1$ and $\sigma_2$ are the widths in Faraday depth space, and $\chi_{0,1}$ and $\chi_{0,2}$ are initial polarization angles of two sources, respectively.

\begin{figure*}
 \begin{center}
  \begin{tabular}{c}
   \begin{minipage}{0.5\hsize}
    \begin{center}
     \includegraphics[width=8cm, bb=0 0 350 250]{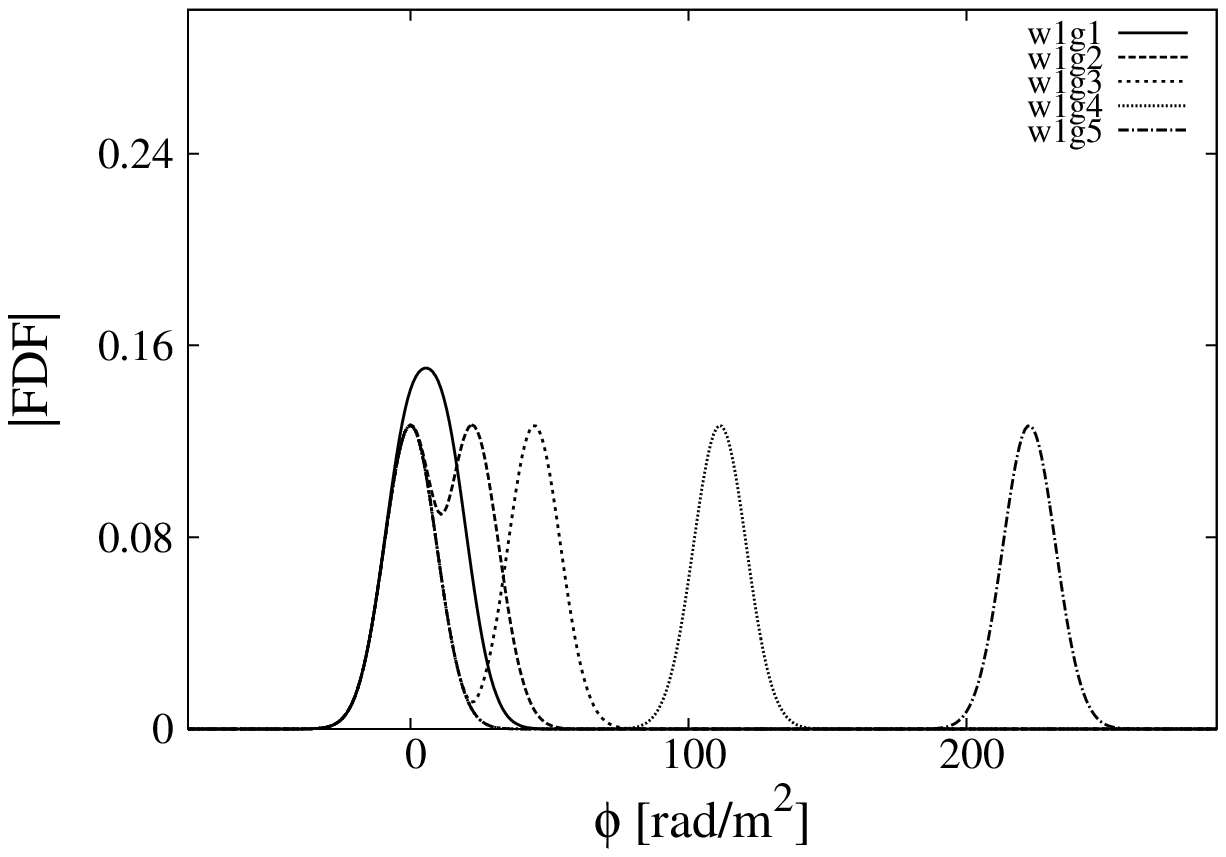} 
    \end{center}
   \end{minipage}
   \begin{minipage}{0.5\hsize} 
    \begin{center}
     \includegraphics[width=8cm, bb=0 0 350 250]{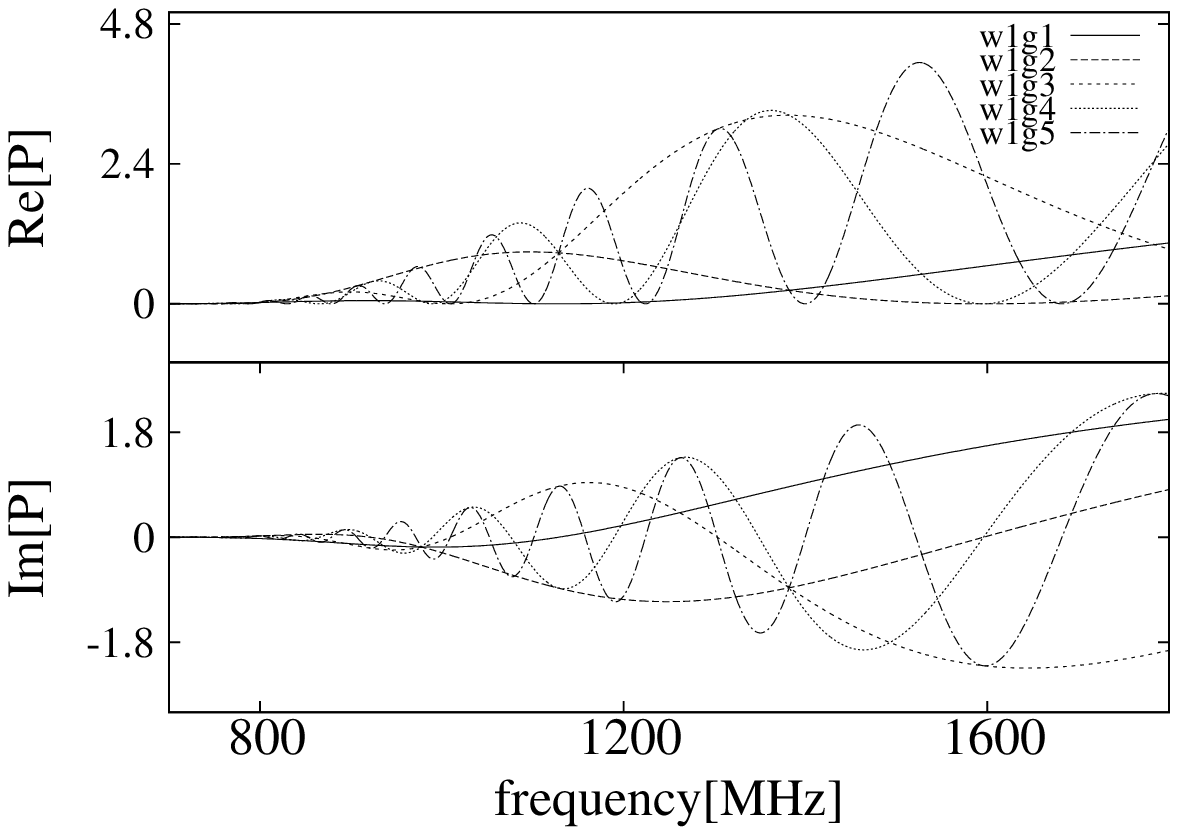} 
    \end{center}
   \end{minipage}
  \end{tabular} 
 \end{center}

 \begin{center}
  \begin{tabular}{c}
   \begin{minipage}{0.5\hsize}
    \begin{center}
     \includegraphics[width=8cm, bb=0 0 350 250]{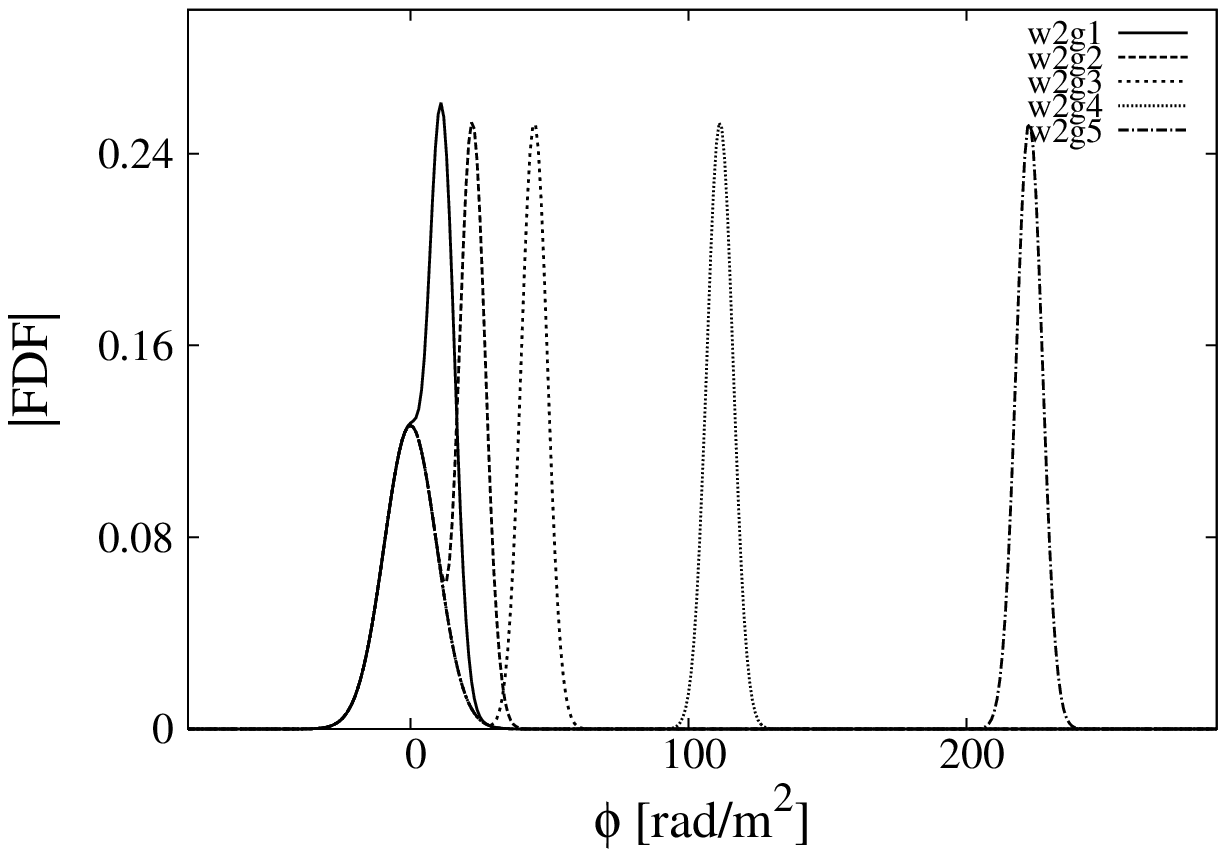} 
    \end{center}
   \end{minipage}
   \begin{minipage}{0.5\hsize} 
    \begin{center}
     \includegraphics[width=8cm, bb=0 0 350 250]{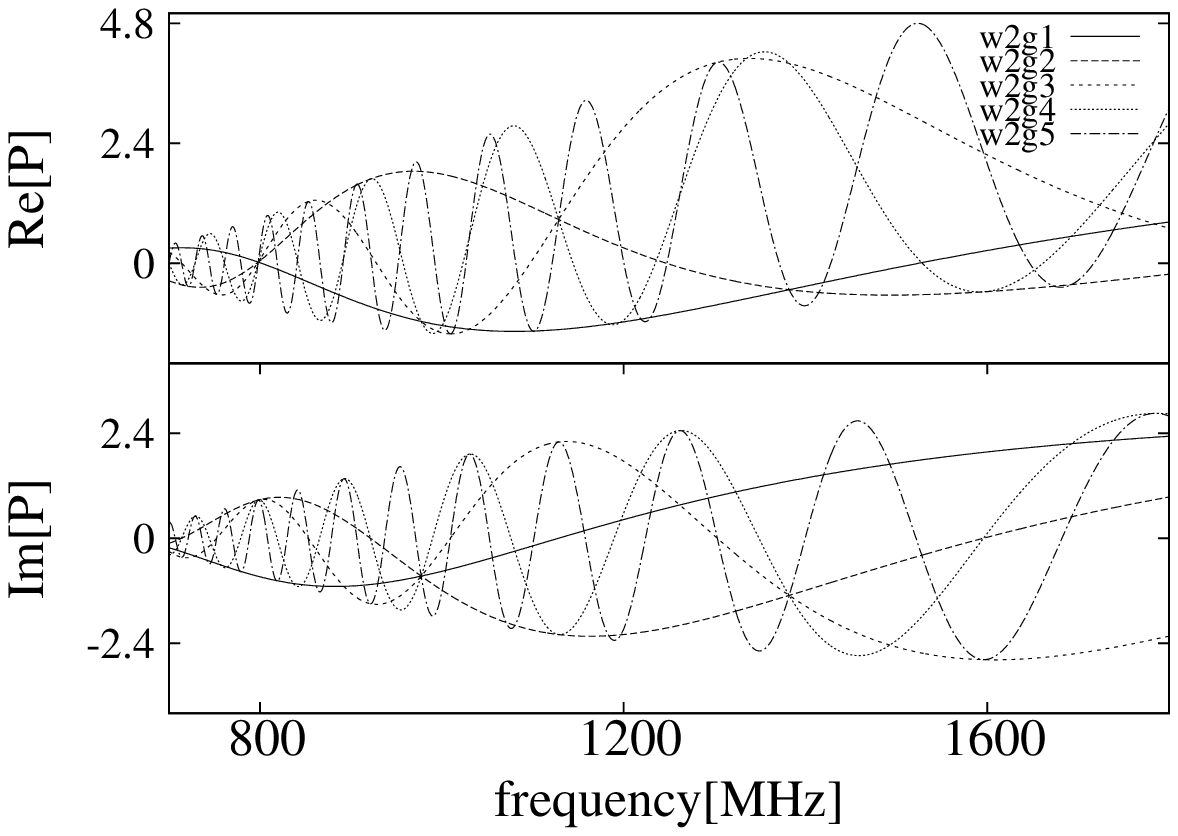} 
    \end{center}
   \end{minipage}
  \end{tabular} 
 \end{center}

 \begin{center}
  \begin{tabular}{c}
   \begin{minipage}{0.5\hsize}
    \begin{center}
     \includegraphics[width=8cm, bb=0 0 350 250]{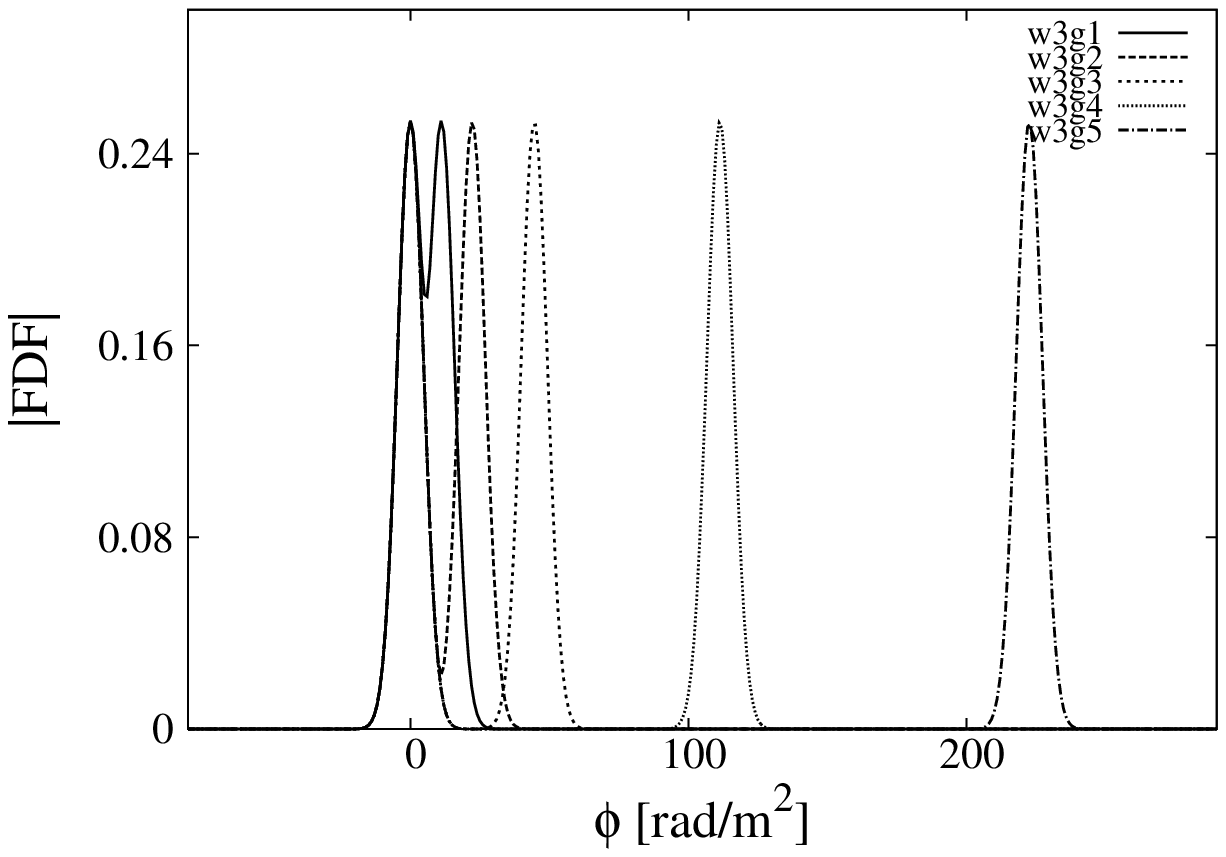} 
    \end{center}
   \end{minipage}
   \begin{minipage}{0.5\hsize} 
    \begin{center}
     \includegraphics[width=8cm, bb=0 0 350 250]{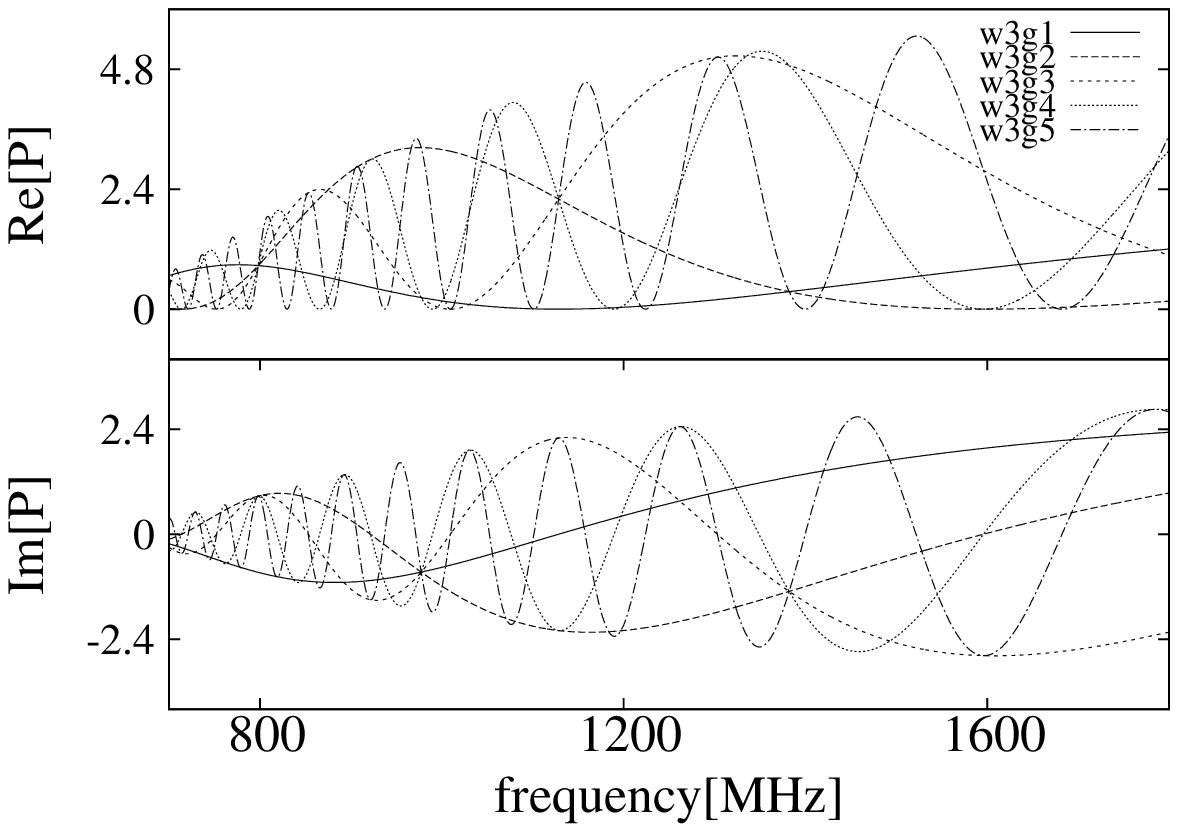} 
    \end{center}
   \end{minipage}
  \end{tabular} 
 \end{center}
\caption{Absolute value of Faraday spectra of source models (left) and the corresponding polarized intensity without noise (right) for w1〜w3 models from top to bottom.}
\label{fig:model}
\end{figure*}

We perform simulations for 15 source models with fixed values of $f_1= 3$, $f_2 = 3$, $\phi_1 = 0$~[rad/m$^2$], $\chi_{0,1} = 0$~[rad], and $\chi_{0,2} = \pi/4$~[rad], and varying the following parameters:
\begin{itemize}
\item {\bf Gap:} The separation of the two Gaussians. $\phi_2 - \phi_1 = \phi_2 =$ 0.5 (g1), 1.0 (g2), 2.0 (g3), 5.0 (g4), and 10.0 (g5) in units of the full width at half maximum (FWHM) of the RMSF.
\item {\bf Width:} The thickness of the two Gaussians. $\sigma_1$ = $\sigma_2$ = 0.5 (w1), $\sigma_1$ = 0.5 and $\sigma_2$ =0.25 (w2), and $\sigma_1$ = $\sigma_2$ = 0.25 (w3) in units of FWHM of the RMSF.
\end{itemize}
We label the source models, for example, w1g3, in the case with $\phi_2 - \phi_1 = 0.5$ FWHM and $\sigma_1 = \sigma_2 = 0.25$ FWHM. The 15 source models are summarized in Table \ref{table:name_model} and plotted in the left panels of Fig.~\ref{fig:model}.

\begin{table}
\caption{A list of source models}
\label{table:name_model}
\scalebox {0.9}{
\begin{tabular}{r|ccc}
\hline \hline
 & Width$^1$ \tnote{1} $\sigma_1$ = $\sigma_2$ & $\sigma_1$ = 0.5 & $\sigma_1$ = $\sigma_2$ \\
 & = 0.5 & $\sigma_2$ = 0.25 & = 0.25 \\
\hline
Gap$^1$ \tnote{1} $\phi_2 - \phi_1$ = 0.5  &  w1g1  &  w2g1  &  w3g1  \\
1  &  w1g2  &  w2g2  &  w3g2  \\
2  &  w1g3  &  w2g3  &  w3g3  \\
5  &  w1g4  &  w2g4  &  w3g4  \\
10  &  w1g5  &  w2g5  &  w3g5  \\
\hline \hline
\end{tabular}
}
\begin{tablenotes}
\item[1]$^1$ In units of the FWHM = 22.26 [rad/m$^2$].
\end{tablenotes}
\end{table}

Using the source FDF models, we make mock polarized intensity spectrum considering the ASKAP full bandwidth (700~MHz -- 1800~MHz, which leads to the FWHM of 22.26 [rad/m$^2$]) with the channel width of 1~MHz. Thus, the number of channels is 1,100 and $n = 2,200$ taking the two polarizations into account. The Stokes Q and U spectra are shown in the right panels of Fig.~\ref{fig:model}. In addition, we add random Gaussian white noise with the mean of 0 and variance of 1 to each channel.

We prepare four fitting models, G1, G2, G3 and G4, which consist of one to four Gaussian function(s), respectively. Because each Gaussian function has four parameters (see Eq.~(\ref{eq:Gauss})), the four models have 4, 8, 12 and 16 parameters, respectively. Physically, the four models have different numbers of polarized sources along the line of sight and, ideally, G2 model should be selected through the process described in the previous subsection.

Next, let us describe the MCMC method we use here. For given mock data and fitting model, we try to find the best-fit parameter set using MCMC where parameter sampling is performed as follows (model parameters are denoted as a vector $\theta$ below).
\begin{itemize}
\item Generate a candidate parameter $\theta'$ from the Gaussian distribution with the previous sample $\theta_t$ as average and the given step width as variance.
\item Accept the candidate and update the parameter by setting $\theta_{t+1} = \theta'$ with the following probability $u$,
\begin{equation}
u = \min\left(1,\frac{L(\theta')}{L(\theta_t)}\right),
\end{equation}
or otherwise the candidate is rejected by setting $\theta_{t+1} = \theta_t$. Here,
\begin{equation}
L(\theta) \propto \exp\left(-\frac{1}{2}\sum_{i=1}^n\left(P_{\rm obs}(\lambda_i^2)-P_{\rm mod}(\lambda_i^2;\theta)\right)^2\right)
\end{equation}
is the likelihood given the parameter vector $\theta$, where $\lambda_i^2$ is the squared wavelength of $i$-th channel and $P_{\rm mod}(\lambda_i^2;\theta)$ is the fitting model calculated with the parameter vector $\theta$.
\end{itemize}
We first run MCMC to tune the step widths of each parameter to ensure the acceptance ratio of $\sim 30\%$. Then we fix the step widths and perform sampling with 20,000 samples. The sample is updated every 1,000 steps and is assessed convergence check using Geweke's diagnostics \citep{geweke}, in which we regard the parameters as being converged if the following condition is satisfied.
\begin{equation}
Z=\frac{\bar{y}_A - \bar{y}_B}{\sqrt{V(y_A)+V(y_B)}}<z,
\end{equation}
where $\bar{y}$ and $V(y)$ are the average and variance of a parameter y in the MCMC chain, respectively, and the subscripts $A$ and $B$ are the first 10\% and latter 50\% sections of the chain, respectively. We adopt $z=1.96$ as the $Z$-value, that corresponds to the significance level of 5\%. The MCMC stops when the chains of all parameters satisfy the convergence condition or the step number reaches the maximum regulation number, which is set to 100,000 in this work.

For each source model, we perform QU-fitting of 4 fitting models to the mock data 100 times with different random noise realizations. Thus, we perform QU-fitting, in total, (15 source models) $\times$ (4 fitting models) $\times$ (100 realizations) $= 6,000$ times.

\section{Results}

\begin{figure}
\includegraphics[width=8cm]{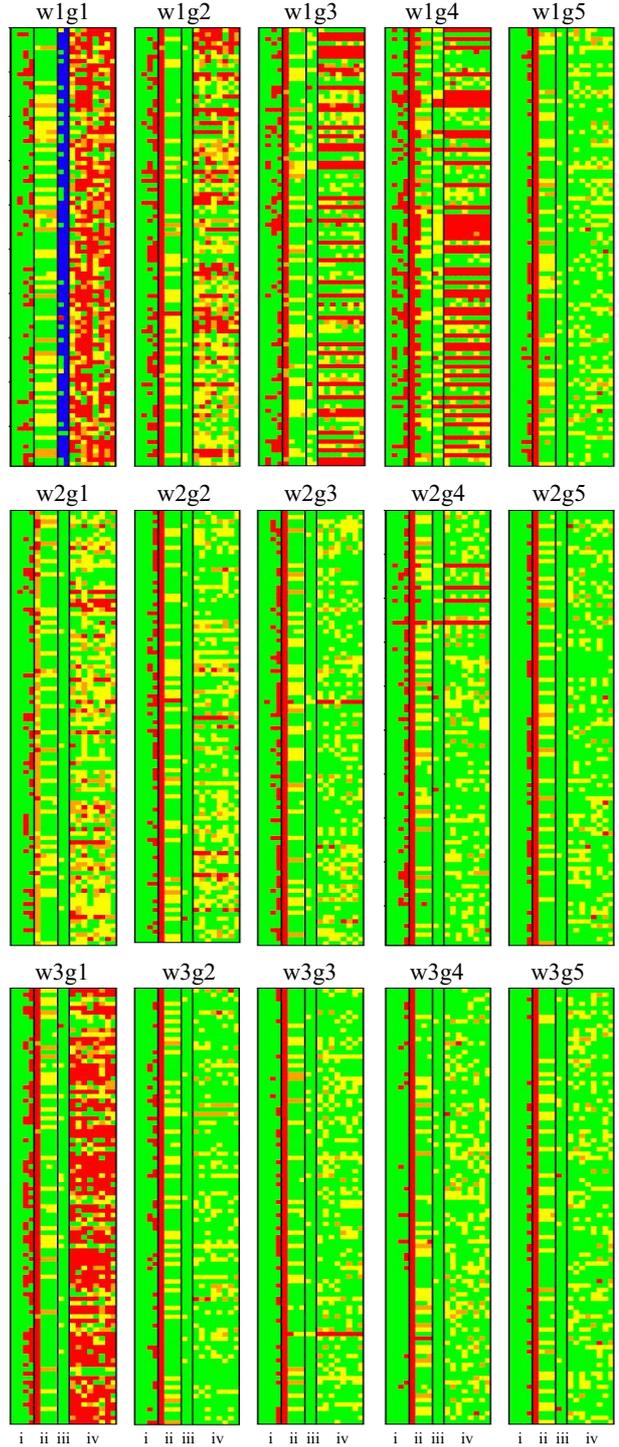}
\caption{
QU-fitting results for 15 source models (see Table~1). The results for 100 realization simulations are shown. Colors are based on our criteria of quality assessment. Criterion (i): MCMC for the G1, G2, G3, and the G4 fitting models are converged (green) or not (red). Criterion (ii): the chi-squared values with the G1, G2, G3, and the G4 fitting models are located within 1$\sigma$ (green), 2$\sigma$ (yellow), 3$\sigma$ (orange), or beyond (red) of the chi-squared distribution. Criterion (iii): AIC/BIC selected G1 (blue), G2 (green), G3 (yellow), or G4 (red). Criterion (iv): the estimated parameters of the G2 model are located in 1$\sigma$ (green), 2$\sigma$ (yellow), 3$\sigma$ (orange), or beyond (red) of the probability distribution.
}
\label{result:sim}
\end{figure}

Fig.~\ref{result:sim} summarizes the results of QU-fitting for the 15 source models. In each panel, the goodness of QU-fitting of 100 realizations are shown from the top to bottom. Each row has 18 colored boxes and represents one realization. Color indicates our criteria of goodness of QU-fitting described as follows.
\begin{itemize}

\item {\bf Criterion (i):} Four boxes from left to right in area (i) show the convergence of MCMC chain with the G1 to the G4 fitting models. Green means that MCMC is converged, and red is not. Note that in red cases, MCMC is quit as reaching the maximum regulation number (100,000).

\item {\bf Criterion (ii):} Four boxes from left to right in area (ii) show the chi-squared values for the G1 to the G4 fitting models, respectively. Green, yellow, orange, and red mean that the chi-squared value is within 1$\sigma$, 2$\sigma$, 3$\sigma$ of the chi-squared distribution with $n$ degrees of freedom, and out of 3$\sigma$, respectively. 

\item {\bf Criterion (iii):} Left and right boxes in area (iii) show the model selection by AIC and BIC, respectively. Blue, green, yellow, and red mean that the G1, G2, G3, and the G4 fitting model was selected by the information criteria, respectively. Note that the source model always consists of two Gaussians and, therefore, the model selection is successful when the box is green.

\item {\bf Criterion (iv):} Eight boxes from left to right in area (iv) show the best-fit values of $\phi_1$, $f_1$, $\chi_{0,1}$, $\sigma_1$, $\phi_2$, $f_2$, $\chi_{0,2}$, and $\sigma_2$ of the G2 fitting model. Green, yellow, orange, and red mean that the true parameter is within 1$\sigma$, 2$\sigma$, 3$\sigma$ of the posterior probability distribution, and out of 3$\sigma$, respectively. Note that the G2 fitting model is not necessarily selected by the information criteria.
\end{itemize}

In Fig.~\ref{result:sim}, we can see global features of the quality of QU-fitting. Generally, as we described above, a panel (a source model) which is mostly green has a high score, that is, QU-fitting is effective for the model. Such models are w1g5, w2g1-5 and w3g2-5.

Let us see the results more closely. The convergence of MCMC (Criterion i) is broadly achieved. The non-convergence appears rather randomly in runs (realizations), meaning that it depends on the observational noise and a condition of parameter search. However no strong correlation is found between MCMC convergence (Criterion i) and goodness of model parameter search (Criterion iv), because parameter estimation is not always good though MCMC chain is converged. The convergence rate depends on both the source model and fitting model. It decreases as the number of model parameters increases from G1 to G4, however it does not have simple correlation between the gap.

Most of the fitting models fit well to the mock data but we obtain large chi-squared values with the G1 fitting model (the leftmost of Criterion ii), indicating that two Gaussian sources cannot be fitted with a single Gaussian function. An exception is w1g1, where two Gaussian components are so close in Faraday depth space that the source model looks like a single Gaussian (the solid line of the top-left panel in Fig.~\ref{fig:model}). Indeed, for w1g1, both AIC and BIC select mostly the G1 fitting model as the best model (Criterion iii), although there are actually two Gaussian sources overlapping each other. For the other source models, except some cases, AIC and BIC tend to choose the G2 fitting model, and rarely select the G3 and the G4 fitting models.

Quality of parameter estimation significantly depends on the source model. We can see that parameter estimation is relatively poor for w1g1, w1g2, w1g3, w1g4, where the widths of two Gaussian sources are relatively large, while, in the case of w1g5 model, estimated values are mostly within 1$\sigma$ since the gap between two Gaussians is significantly larger than the widths. The source model of w3g1 also leads to poor estimation, although the model has a double-peak profile as seen in Fig.~\ref{fig:model}. QU-fitting tends to show a good reconstruction in the cases that the source models have relatively smaller widths, a larger gap, and also the different shape (all of w2 models). We can understand from the right panels of Fig.~\ref{fig:model} that smaller widths lead to better results because depolarization is less significant for smaller widths.

Next let us closely see the results of the G2 fitting model. We count the success rates for the G2 fitting model on (i) convergence of MCMC chain, (ii) the chi-squared value within 3$\sigma$ of the chi-squared distribution, (iii) selection by AIC/BIC, and (iv) all true parameters within 3$\sigma$ in their posterior probability distributions. It should be noted that the above success rates are counted irrespective of the selected model.

\begin{table}
  \caption{The success rate (\%) of QU-fit for G2 fitting model.}
  \label{table:evaluation}
  \centering
  \begin{tabular}{l|ccccc}
    \hline
   &  (i)  & (ii) & (iii)AIC & (iii)BIC & (iv) \\
   \hline \hline
   w1g1 & 93 & 100& 26 & 1 & 1  \\
   w1g2 & 92 & 99 & 91 & 100 & 50  \\
   w1g3 & 85 & 100 & 62 & 91 & 58  \\
   w1g4 & 62 & 57 & 54 & 56 & 52   \\
   w1g5 & 100 & 100 & 89 & 100 & 96 \\
   w2g1 & 99 & 100 & 93 & 100 & 71   \\
   w2g2 & 100 & 99 & 91 & 100 & 91  \\
   w2g3 & 99 & 99 & 89 & 99 & 97  \\
   w2g4 & 95 & 96 & 88 & 96 & 95   \\
   w2g5 & 100 & 100 & 96 & 100 & 94  \\
   w3g1 & 100 & 100 & 88 & 100 & 17   \\
   w3g2 & 100 & 100 & 93 & 100 & 97  \\
   w3g3 & 100 & 99 & 90 & 99 & 99  \\
   w3g4 & 100 & 99 & 91 & 100 & 97  \\
   w3g5 & 100 & 100 & 91 & 100 & 98  \\
	\hline
  \end{tabular}
\end{table}

Table~\ref{table:evaluation} shows the success rates. We find that for w1g5, w2g1--w2g5, and w3g2--w3g5, QU-fitting is fairly successful in all aspects (i) -- (iv). This is because these models have a sufficient gap between the two sources to separate them. It is interesting to notice that QU-fitting works quite well for w2g1 even though the two sources overlap each other (see Fig.\ref{fig:model}). On the other hand, the other models show noticeably low success rates on (iii) model selection and (iv) parameter estimation. In the case of w1g1, the model selection is very poor: AIC and BIC select the G1 fitting model by 70\% and 99\%, respectively, because w1g1 source model has only one peak as already mentioned. Model selection is relatively poor for w1g3 and w1g4 as well. In the cases of w1g2 and w3g1, AIC and BIC select the correct the G2 fitting model by about 90\% or more, while parameter estimation is successful only by about 50\% and 20\%, respectively.

Comparing AIC and BIC, BIC generally has slightly higher scores except w1g1, where FDF looks like a single source. Because the sum of two Gaussian functions cannot be expressed as a single Gaussian, fitting w1g1 mock data with the G1 fitting model cannot be perfect. Nevertheless, because the penalty imposed by BIC is relatively large, it will tend to select the G1 model.

\begin{figure}
\begin{center}
\includegraphics[width=8cm]{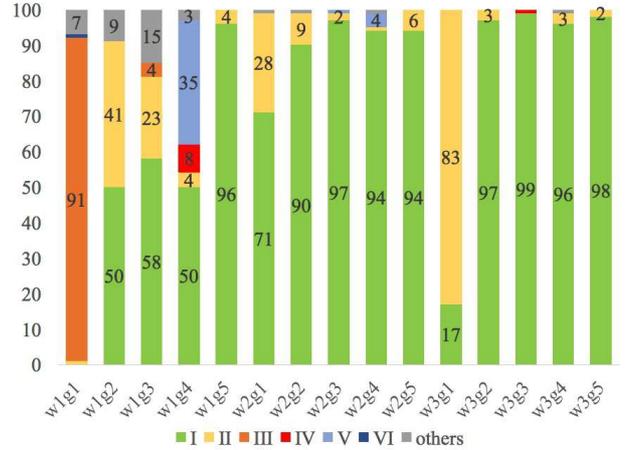}
\end{center}
\caption{
Results of the categorization for achievement of model fitting.
}
\label{table:tendency_all}
\end{figure}

In order to clarify at which step the QU-fitting fails to reproduce the source model, we define a categorization of the results in all of possible situation in Table~\ref{table:possible_result}. Fig.~\ref{table:tendency_all} shows the breakdown of the results into the categorization according to Table~\ref{table:possible_result}. Again, w1g5, w2g1--w2g5, and w3g2--w3g5 are very successful at all stages, that is, classified mostly into case I (green), while w1g1 fails at (iii) model selection (orange), w1g2, w1g3 and w3g1 fail at (iv) parameter estimation (yellow), and w1g4 fails from (i) MCMC convergence (red). 

\begin{table}
  \caption{Categorization of the results} 
  \label{table:possible_result}
  \centering
  \begin{tabular}{l|cccc}
    \hline
     &  (i)  & (ii) & (iii)BIC  & (iv)   \\
   \hline \hline
   case I    &  $\circ$  &  $\circ$  &  $\circ$  &  $\circ$   \\
   case II   &  $\circ$  &  $\circ$  &  $\circ$  &  $\times$  \\
   case III  &  $\circ$  &  $\circ$  &  $\times$ &  $\times$  \\
   case IV   &  $\circ$  &  $\times$ &  $\times$ &  $\times$  \\
   case V    &  $\times$ &  $\times$ &  $\times$ &  $\times$  \\
   case VI   &  $\circ$  &  $\circ$  &  $\times$ &  $\circ$   \\
   others & - & - & - & - \\
    \hline
  \end{tabular}
\end{table}

\section{Discussion}

In this section, we investigate source models with which QU-fitting does not work well, w1g1--w1g4 and w3g1, and try to clarify the tendencies of the results. Overall, these models are the cases with relatively thick Faraday spectra and/or a relatively narrow gap between two components.

\subsection{Local-Maximum Trapping}

In cases II--V of Table~\ref{table:possible_result}, MCMC chain fails to reach the true parameter values. One of the possible reasons is that the chain is trapped in a local maximum of the likelihood function in parameter space. Because we set the initial values of all parameters to 0 as is often done, trapping in local maxima can often happen depending on the structure of the likelihood function. In order to check this possibility, we perform simulations again, with the initial parameters set to the true values.

\begin{table}
  \caption{Success rates (\%) of model selection and parameter estimation with/without true initial values. The former is indicated by "+" in the first row and the latter is the same as Table~\ref{table:evaluation}.}
  \label{table:evaluation_ans}
  \centering
  \small
  \scalebox{0.85}{
  \begin{tabular}{l|cccccc}
    \hline
   & (iii)AIC & (iii)AIC+ & (iii)BIC & (iii)BIC+ & (iv) & (iv)+ \\
   \hline \hline
   w1g1 & 26 &  22 &   1 &   1 &  1 &  3 \\
   w1g2 & 91 &  89 & 100 & 100 & 50 & 50 \\
   w1g3 & 62 &  86 &  91 & 100 & 58 & 92 \\
   w1g4 & 54 &  96 &  56 & 100 & 52 & 98 \\
   w3g1 & 88 &  90 & 100 & 100 & 17 & 45 \\
    \hline
  \end{tabular}
  }
\end{table}

Table~\ref{table:evaluation_ans} shows the success rates of model selection and parameter estimation for w1g1, w1g2, w1g3, w1g4 and w3g1, with which the rates of (iii) and/or (iv) are relatively low ($< 80\%$, see Table~\ref{table:possible_result}). We can see that starting with true parameter values significantly improves the success rates of both model selection and parameter estimation for w1g3 and w1g4 models, indicating that the local-maximum trapping is a main cause of failure. On the other hand, this is not the case with w1g1 and w1g2. For w3g1, although some improvement can be seen, the success rate (iv) is still low. It will be essentially difficult for QU-fitting to resolve multiple components overlapping with each other significantly.

Although true initial values are unknown in real observations, RM CLEAN can be used to estimate the number of sources and rough values of parameters. If we use the estimated parameters as initial values of the parameters, QU-fitting will be more effective.

\subsection{Fake Source}

\begin{figure}
\begin{center}
\includegraphics[width=8cm]{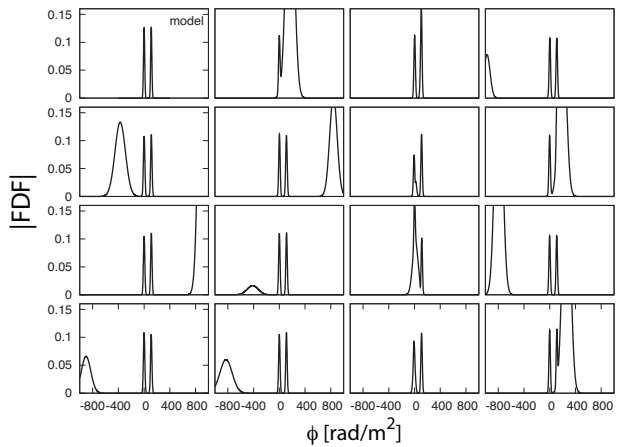}
\end{center}
\caption{
Reconstructed Faraday spectra with G3 fitting models for w1g4. The top left panel show the w1g4 source model ($\phi_1$ = 0 [rad/m$^2$], $\phi_2$ = 111.3 [rad/m$^2$]).
}
\label{fdf_w1g4}
\end{figure}

QU-fitting sometimes overestimates the number of components by selecting the G3 fitting model. In particular, for w1g4, the G3 model is selected 38 times out of 100 realizations by BIC. Among the 38 cases, 30 realizations are classified into case V, where fitting by the G2 model failed completely. Thus, when the G2 model fails to fit, the G3 model tends to be selected. When fitting is performed with G3 model, it often happens that two of the components reproduce the true components correctly, while the last one is located at large absolute Faraday depth with a large thickness.

Fig.~\ref{fdf_w1g4} shows the reconstructed Faraday spectra of 15 examples out of the 38 realizations which select the G3 fitting model for w1g4. As we can see, there is a Faraday-thick component at large absolute Faraday depth. In fact, the extra component does not significantly contribute to polarized intensity in the frequency range of ASKAP because of strong depolarization effect. Therefore, we expect that such an extra component with a width larger than the max-scale (the largest scale in $\phi$ space to which one is sensitive) can be recognized as a fake, and then removed.

\section{Summary}

In this paper, we examined the functionality of standard QU-fitting algorithm quantitatively by simulating spectro-polarimetric observations of two extent sources located along the same LOS. We assumed the Gaussian function as a model of the extent sources and varied the gap and width of the two Gaussians systematically. Especially, focusing on the convergence of MCMC chain, obtained chi-squared value, model selection by AIC and BIC, and parameter estimation, we evaluated the effectiveness of QU-fitting. For source models with a sufficient gap between two sources, QU-fitting works fairly well. Contrastingly, overlapping thick sources are difficult to be separated (w1g1), while overlapping thin sources can be separated (w2g1 and w3g1). Further, even if two sources do not overlap with each other, model selection and/or parameter estimation often fail for sources as thick as the FWHM determined by the observation frequency band (w1g2, w1g3 and w1g4). This is partly due to the trapping of MCMC chain to a local maximum of likelihood function. So, if we could obtain rough estimate of parameter values from RM CLEAN, for example, QU-fitting works even better by setting the initial values of MCMC chain to the estimated values. We considered these four criteria of the quality of QU-fitting independently. However, in fact, they are closely related to each other. For example, if the obtained chi-squared value is large even though the MCMC chain is converged, it is interpreted that the chain was trapped in a local maximum of likelihood function. In this case, model selection and parameter estimation would also fail. To avoid this type of failure, more sophisticated algorithm of MCMC will be necessary. Another possible example is a case where the MCMC chain is converged and a reasonable chi-squared value is obtained, but model selection is failed. This example would imply the limitation of the information criterion.

\section*{Acknowledgements}

This work is supported in part by Grand-in-Aid from the Ministry of Education, Culture, Sports, and Science and Technology (MEXT) of Japan, No. 24340048 (KT), 26610048 (KT), 15H03639 (TA), 15H05896 (KT), 15K17614 (TA), 16H05999 (KT), and 17H01110 (TA, KT), Bilateral Joint Research Projects of JSPS (KT), and by the National Research Foundation of Korea through grant 2007-0093860 (SI).












\bsp	
\label{lastpage}
\end{document}